\documentclass[12pt,preprint]{aastex}

\shorttitle{Collapsed Cores in Globular Clusters}
\shortauthors{Djorgovski et al.}

\title{Cosmological-model-independent tests for the distance-duality relation from Galaxy Clusters and Type Ia Supernova}

\author{Zhengxiang Li$^{a}$, Puxun Wu$^{b}$  and Hongwei Yu$^{a, b,} \footnote{Corresponding author: hwyu@hunnu.edu.cn}$ }

\affil{$^a$Institute of Physics and Key Laboratory of Low
Dimensional Quantum Structures and Quantum Control of Ministry of
Education, Hunan Normal University, Changsha, Hunan 410081, China
\\$^b$Center for Nonlinear Science and Department of Physics, Ningbo
University,  Ningbo, Zhejiang 315211, China }

\begin{abstract}

We perform a cosmological-model-independent  test for the
distance-duality (DD) relation $\eta(z)=D_L(z)(1+z)^{-2}/D_A(z)$,
where $D_L$ and  $D_A$ are the luminosity distance and  angular
diameter distance respectively, with a combination of observational
data for $D_L$ taken from  the latest Union2 SNe Ia and that for
$D_A$ provided by two galaxy clusters samples compiled by De
Filippis {\it et al.} and Bonamente {\it et al.}. Two
parameterizations for $\eta(z)$, i.e.,  $\eta(z)=1+\eta_0z$ and
$\eta(z)=1+\eta_0z/(1+z)$, are used. We find that the DD relation
can be accommodated at $1\sigma$ confidence level (CL) for the De
Filippis {\it et al.} sample and at $3\sigma$ CL for the Bonamente
{\it et al.} sample. We also examine the DD relation by postulating
two more general parameterizations: $\eta(z)=\eta_0+\eta_1z$ and
$\eta(z)=\eta_0+\eta_1z/(1+z)$,  and find that the DD relation is
compatible with the results from the De Filippis {\it et al.} and
the Bonamente {\it et al.} samples at $1\sigma$ and $2\sigma$ CLs,
respectively. Thus, we conclude that the DD relation is compatible
with present observations.

\end{abstract}


\begin{document}

\maketitle 

\section{INTRODUCTION}
The distance-duality (DD) relation~\citep{Etherington} between the
luminosity distance $D_L$ and the angular diameter distance (ADD)
$D_A$, i.e.,
\begin{equation} \frac{D_L}{D_A}(1+z)^{-2}=1,
\end{equation}where $z$ is the redshift,
 plays an important role in modern observational
cosmology~\citep{Schneider, Cunha, Mantz, Komatsu}, and, actually,
it has heretofore been applied to all analysis of the cosmological
observations without any doubt. However, in reality, it is possible
that  one of the requirements in obtaining the DD relation may be
violated. A violation of the DD relation may even be considered as a signal of
the breakdown of physics on which the DD relation is based upon
\citep{Csaki, Bassetta,Bassettb}).

Thus, it is desirable to perform a validity check on  the DD
relation by the astronomical observations. In this regard,
\citet{Uzan} have tested it by using the observations from the
Sunyaev-Zeldovich effect (SZE) and X-ray surface brightness from
galaxy clusters, and found that the DD relation is consistent with
observations at $1\sigma$ confidence level (CL). With a different
galaxy clusters sample provided by \citet{Bonamente},
\citet{Bernardis} also obtained a non-violation of the DD relation.
 In addition, by combining the Union Type Ia supernovae (SNe
Ia) \citep{Kowalski} with the latest measurement of the Hubble
expansion at redshifts between $0$ and $2$ \citep{Stern},
\citet{Avgoustidis} discussed this relation and obtained that it is
consistent with observations at $2\sigma$ CL.  Recently, by assuming
that the DD relation satisfies the following expression
\begin{equation}
\frac{D_L}{D_A}(1+z)^{-2}=\eta(z),
\end{equation}
where $\eta(z)$ is parameterized as  $\eta(z)=1+\eta_0z$ and
$\eta(z)=1+\eta_0z/(1+z)$, \citet{Holandaa} discussed the validity
of the DD relation with the  ADD $D_A$  measurements from galaxy
clusters provided by the \citet{Filippis} (elliptical $\beta$ model)
sample and the \citet{Bonamente} (spherical $\beta$ model) sample,
and the luminosity distance $D_L$ given in the context of
$\Lambda$CDM. Here, the elliptical and spherical $\beta$ models are
two different geometries  used to describe the galaxy clusters.
Their results showed that the elliptical model is more compatible
with no violation of the DD relation. However, all the
aforementioned analyses are model dependent since a cosmic
concordance model ($\Lambda$CDM) is assumed in their discussions. It
is worth noting that \citet{Bernardis} have in fact tried to  test
the DD relation in a  model-independent way.  In their method, the
ADD is given from galaxy clusters and the luminosity distance is
from SNe Ia. To obtain the values of the ADD and the luminosity
distances at the same redshift, \citet{Bernardis} binned their data
and found that the DD relation is not violated at $1\sigma$ CL.
However,
when they determine the ADD from observations, the relation
$D_A^{cluster}(z)=D_A(z)$ is used, which holds  under the condition
with no violation of the DD relation.

Recently, a consistent cosmological-model-independent test for the
reciprocity relation was proposed by \citet{Holandab}. The main idea
 is to test the DD relation directly with the
observed luminosity and ADDs, provided by SNe Ia and galaxy clusters
samples, respectively.   They considered two specific different
redshift-dependent parameterizations for $\eta(z)$:
$\eta(z)=1+\eta_0z$ and $\eta(z)=1+\eta_0z/(1+z)$. The data sets
used  were given from the Constitution SNe Ia \citep{Hicken} and two
ADD samples \citep{Filippis, Bonamente} from galaxy clusters
obtained through SZE effect and X-ray measurement with different
geometry descriptions to the cluster: the elliptical $\beta$ model
\citep{Filippis} and the spherical $\beta$ model \citep{Bonamente}.
They found that the result from the elliptical model is consistent
with the DD relation at $2\sigma$ CL, while for the spherical model
the relation is clearly incompatible with observations.

However,  we find that six and twelve  ADD data points should  be
removed, respectively, for the De Filippis et al. and Bonamente et
al. samples, instead of only three ADD data points that were
discarded for both samples in {Holanda} {\it et al.} (2010b). So, in
the present Letter, we first redo the same analysis as
\citet{Holandab} but with more data points removed to see how this
would affect the result, and we obtain  a more serious violation of
the DD relation. We then consider the effect of the errors of SNe Ia
which were neglected in \citet{Holandab} and give a comparison of
the results obtained with and without the errors of SNe Ia. More
importantly, we retest the DD relation using the latest Union2 SN Ia
data \citep{Amanullah}.

Compared with the Constitution  set used by \citet{Holandab}, the
Union2 has the following advantages: (1) the selection criteria
($\Delta z <0.005$) can be satisfied for all data points of two ADD
samples except for the cluster CL J1226.9+3332 ($z=0.890$), which
corresponds to $\Delta z =0.005$, from the Bonamente et al. (2006)
sample, (2) the values of $z_{SNe~Ia}-z_{Cluster}$ are more centered
around the $\Delta z=0$ line. So, the results from Union2 may be
more reliable. Finally, we test the DD relation by assuming two more
general parameterizations: $\eta(z)=\eta_0+\eta_1z$ and
$\eta(z)=\eta_0+\eta_1z/(1+z)$, and find that in this case the
consistencies between the observations and the DD relation are
improved markedly for both  samples of galaxy clusters.

\section{DATA AND ANALYSIS METHOD}
To obtain  constraints on free parameters in the  parameterizations
of $\eta(z)$: $\eta(z)=1+\eta_0z$  and $\eta(z)=1+\eta_0 z/(1+z)$,
we first  need to  get  $\eta_{obs}$, which can be determined by the
observed luminosity distance $D_L$ and ADD $D_A$ at the same
redshift. The observed ADD $D_A$ is given by the galaxy clusters
obtained by combining the SZE+X-ray surface brightness measurements.
It must be emphasized that,  if a redshift-dependent expression for
the DD relation is considered, the SZE+X-ray surface brightness
observation technique gives
$D_A^{\mathrm{cluster}}(z)=D_A(z)\eta^2(z)$ \citep{Sunyaev,
Cavaliere}. So, we must replace the ADD $D_A(z)$ with
$D_A^{\mathrm{cluster}}(z)\eta^{-2}$ when we try to test the
reciprocity relation  consistently with the SZE+X-ray observations
from galaxy clusters. Thus, the observed $\eta_{obs}(z)$ is
determined by the following expression:
\begin{eqnarray}\eta_{obs}(z)=D_A^{\mathrm{cluster}}(z)(1+z)^2/D_L(z).\end{eqnarray}

For $D_A^{\mathrm{cluster}}$, we consider two samples.  The first
one, including a selection of 18 galaxy clusters in the redshift
interval $0.14<z<0.8$ compiled by \citet{Reese}  and a sample of 7
clusters compiled by \citet{Mason}, was studied and corrected by
\citet{Filippis} with an isothermal elliptical $\beta$ model. The
second is the \citet{Bonamente} sample. It consists of 38 ADD galaxy
clusters  analyzed by assuming the hydrostatic equilibrium model and
spherical geometry for the cluster plasma and dark matter
distributions (spherical $\beta$ model).

\begin{figure}[h!]
   \centering
       \includegraphics[width=0.45\linewidth]{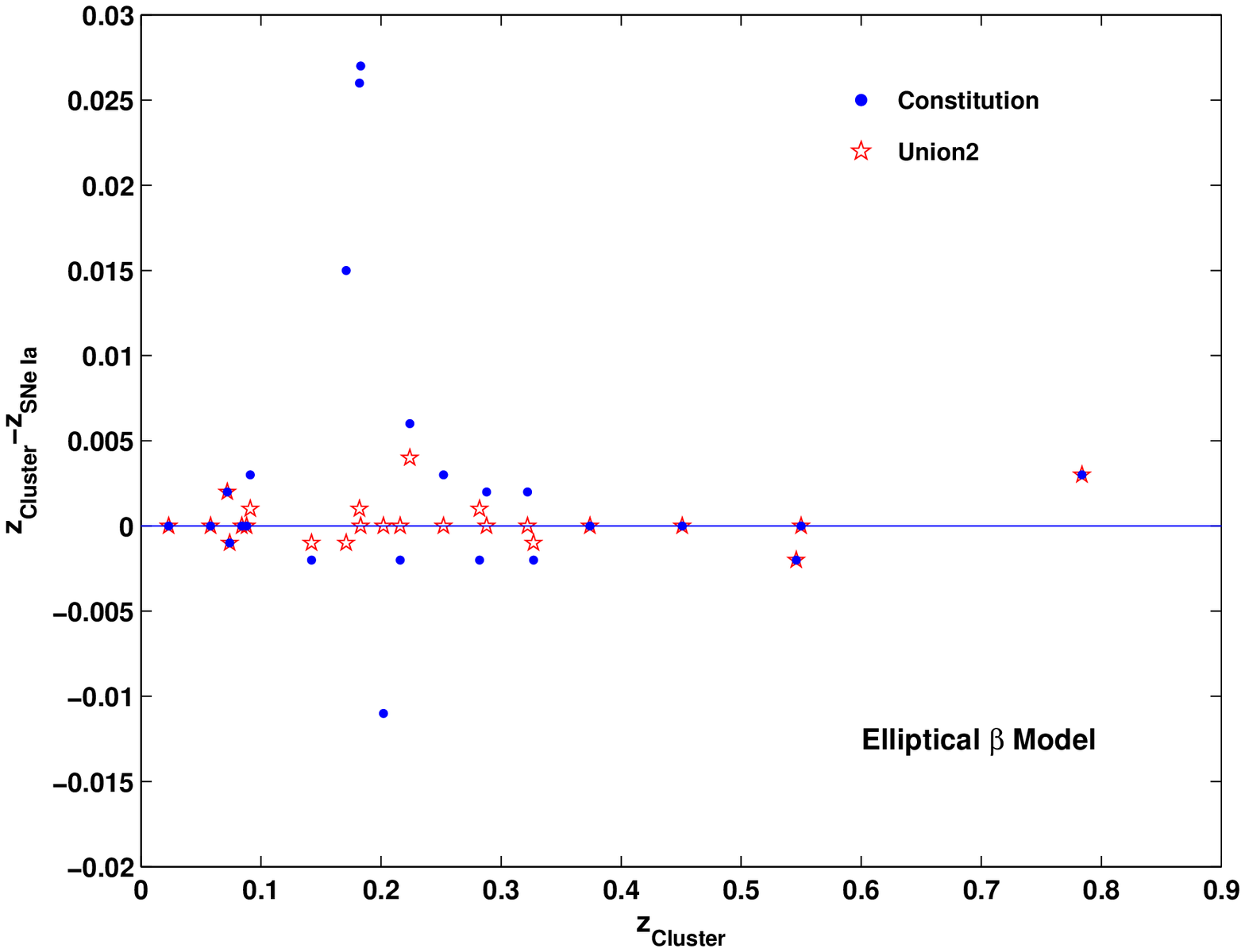}
       \includegraphics[width=0.45\linewidth]{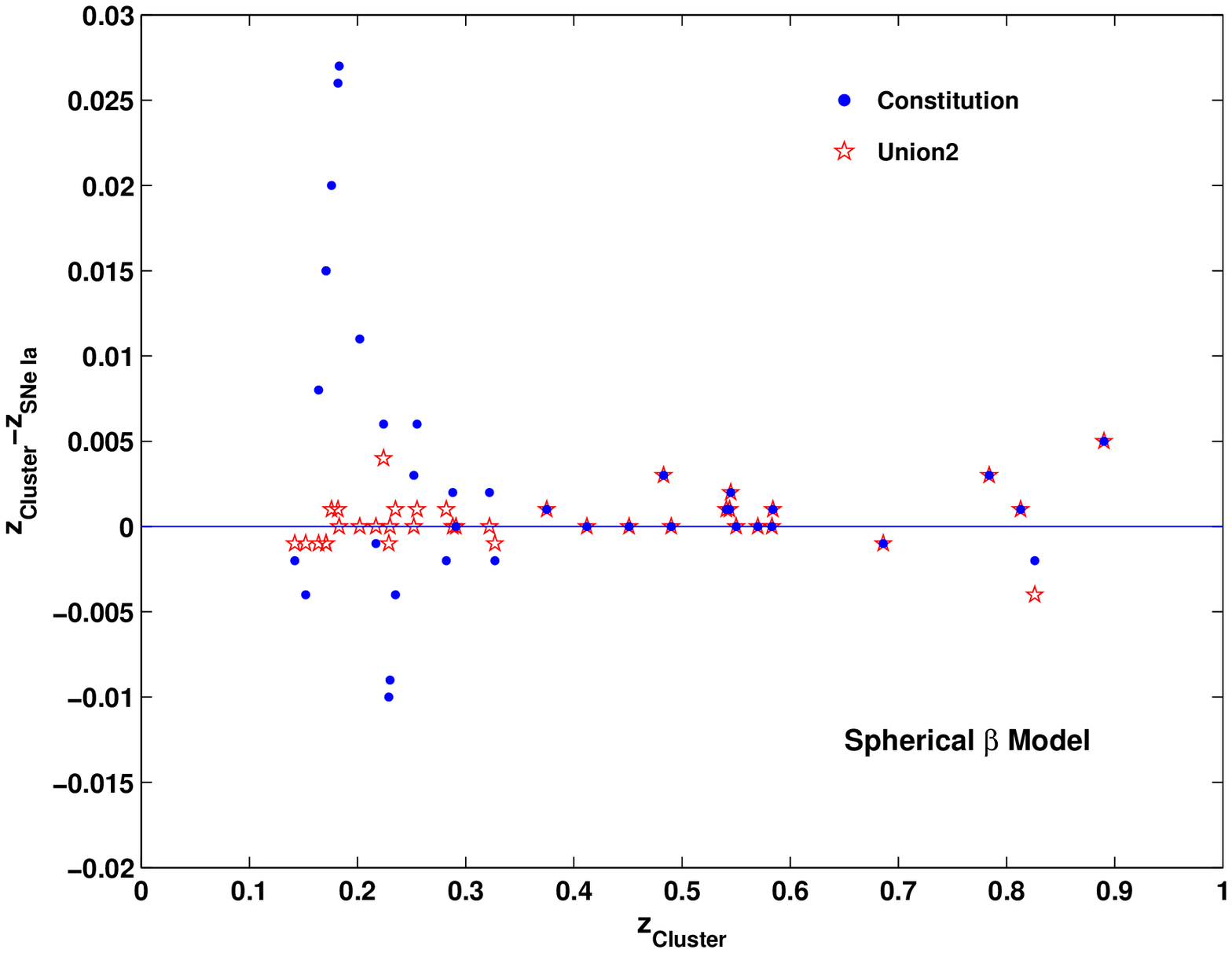}
   \caption{\label{Fig1} Redshift subtraction for the cluster-SN Ia samples.
   The left panel shows the results from De Filippis et al. (2005) sample, while
   the right panel from the Bonamente et al. (2006) sample.
   The blue points and red pentagrams represent the Constitution and Union2 data
   respectively.}
\end{figure}

 For the luminosity distance $D_L$, both the
Constitution and Union2 SN Ia data sets are considered.  For a given
$D_A^{{cluster}}$ data point, theoretically,  we should select an
associated SNe Ia data point at the same redshift to obtain an
$\eta_{obs}$. However, in reality,  it is almost impossible to have
both $D_A^{{cluster}}$ and $D_L$ at exactly the same redshift. So,
as \citet{Holandab}, we use the criteria ($\Delta z=
\left|z_{{Cluster}}-z_{{SNe~Ia}}\right|<0.005$) to select the SNe Ia
data. For Constitution SNe Ia,  we find that there exist six data
points (A2261, A2163, A520, A1689, A665, A2218) for the
\citet{Filippis} sample (elliptical $\beta$ model) and $12$ data
points (A68, A267, A586, A665, CL J1226.9+3332, A1689, A1914, A2111,
A2163, A2218, A2259, A2261) for the \citet{Bonamente} sample
(spherical $\beta$ model) that have to be discarded  (listed in
detail in Fig.~(\ref{Fig1})). This is, however, differs from what is
obtained by \citet{Holandab}, where
 only three data points are removed for both ADD samples.

To get more reliable results, we use the latest Union2 SNe Ia
\citep{Amanullah}  in our analysis.
 The main advantages of the Union2, as compared with the Constitution,
 are: (1) the selection criteria ($\Delta z<0.005$) can be
satisfied for all data points of both ADD samples except for the
cluster CL J1226.9+3332 ($z=0.890$) from the \citet{Bonamente}
sample, which gives $\Delta z=0.005$, (2) points
$z_{SNe~Ia}-z_{Cluster}$ are more centered around the line $\Delta
z=0$, as shown in Fig.~(\ref{Fig1}) which plot the subtractions of
redshifts between the clusters and the associated SNe Ia. Thus, with
Union2, the accuracy of our test will be improved. In order to
ensure the integrity of the ADD samples, we keep the cluster CL
J1226.9+3332 ($\Delta z=0.005$) in our analysis.

Using these observational  data of galaxy clusters and SNe Ia and
the selection criteria ($\Delta z=
\left|z_{{Cluster}}-z_{{SNe~Ia}}\right|<0.005$), we can obtain the
$\eta_{obs}$. To estimate the model parameters of a given
parameterized form, we use the minimum $\chi^2$ estimator following
standard route
\begin{equation}
\chi^2(z;\mathbf{p})=\sum_z\frac{[\eta(z;\mathbf{p})-\eta_{obs}(z)]^2}{\sigma^2_{\eta_{obs}}},
\end{equation}
where
  $\sigma_{\eta_{obs}}$ is the error of $\eta_{obs}$ associated
with the observational technique, and $\mathbf{p}$ represents the
model parameter set.  $\sigma_{\eta_{obs}}$  comes from the statical
contributions and systematic uncertainties of the galaxy clusters
which  are combined in quadrature \citep{DAgostini} and those of SNe
Ia. In our analysis, we consider two cases, i.e., those with and
without the errors of SNe Ia, and then show  the effect of the
errors from SNe Ia which was neglected by \citet{Holandab}

\section{RESULTS}

\begin{figure}[h!]
   \centering
       \includegraphics[width=0.45\linewidth]{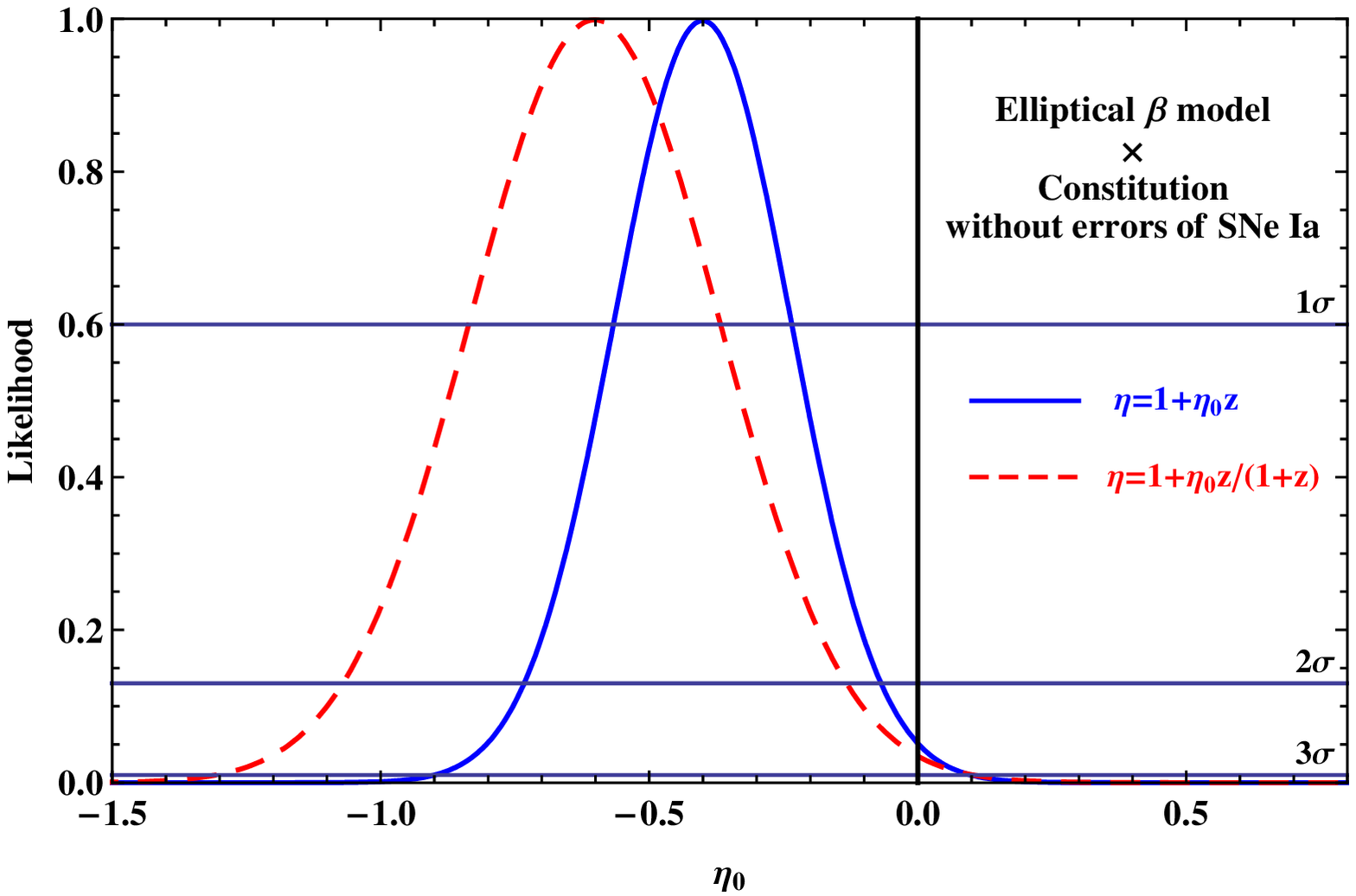}
       \includegraphics[width=0.45\linewidth]{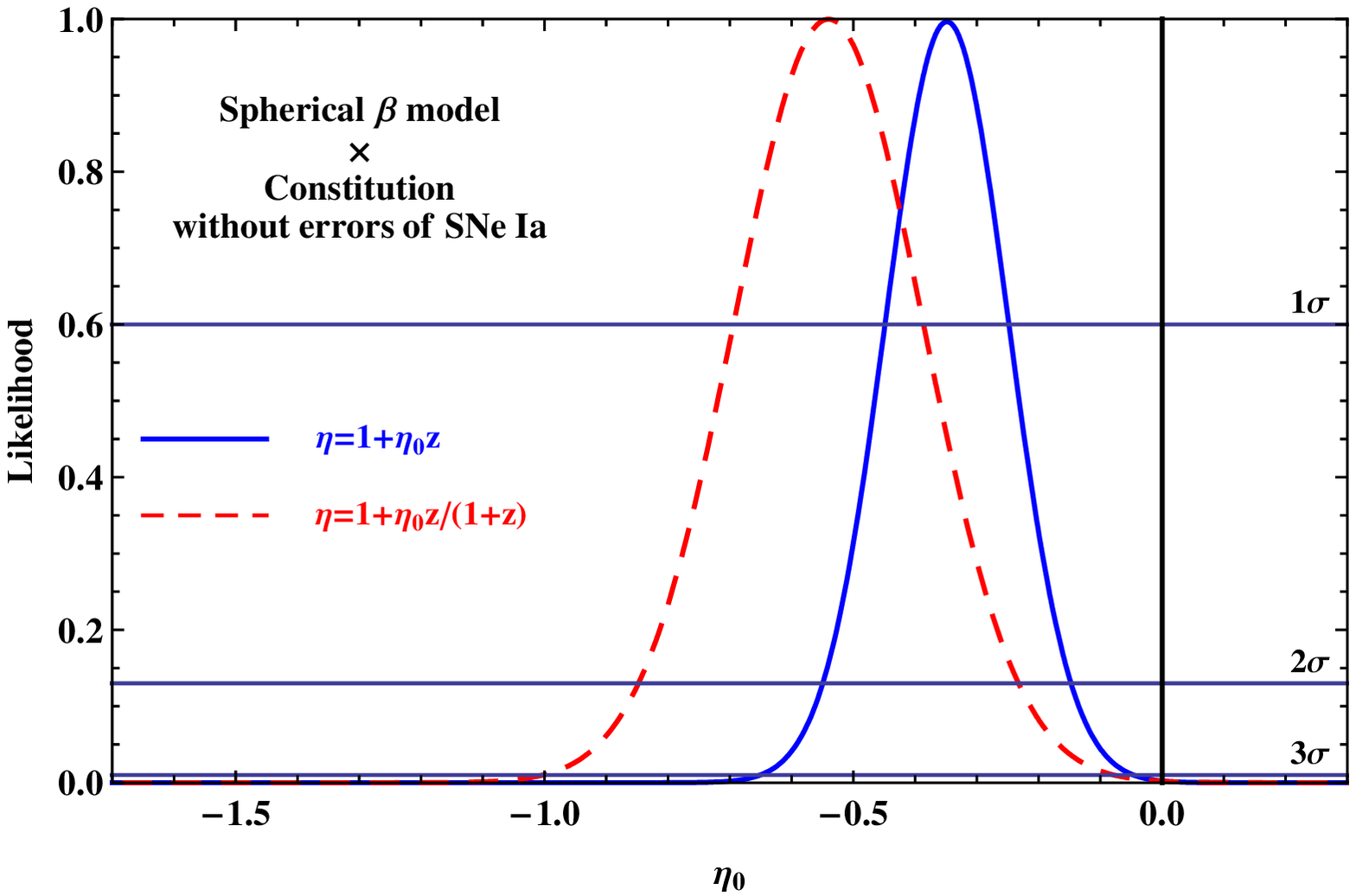}
       \includegraphics[width=0.45\linewidth]{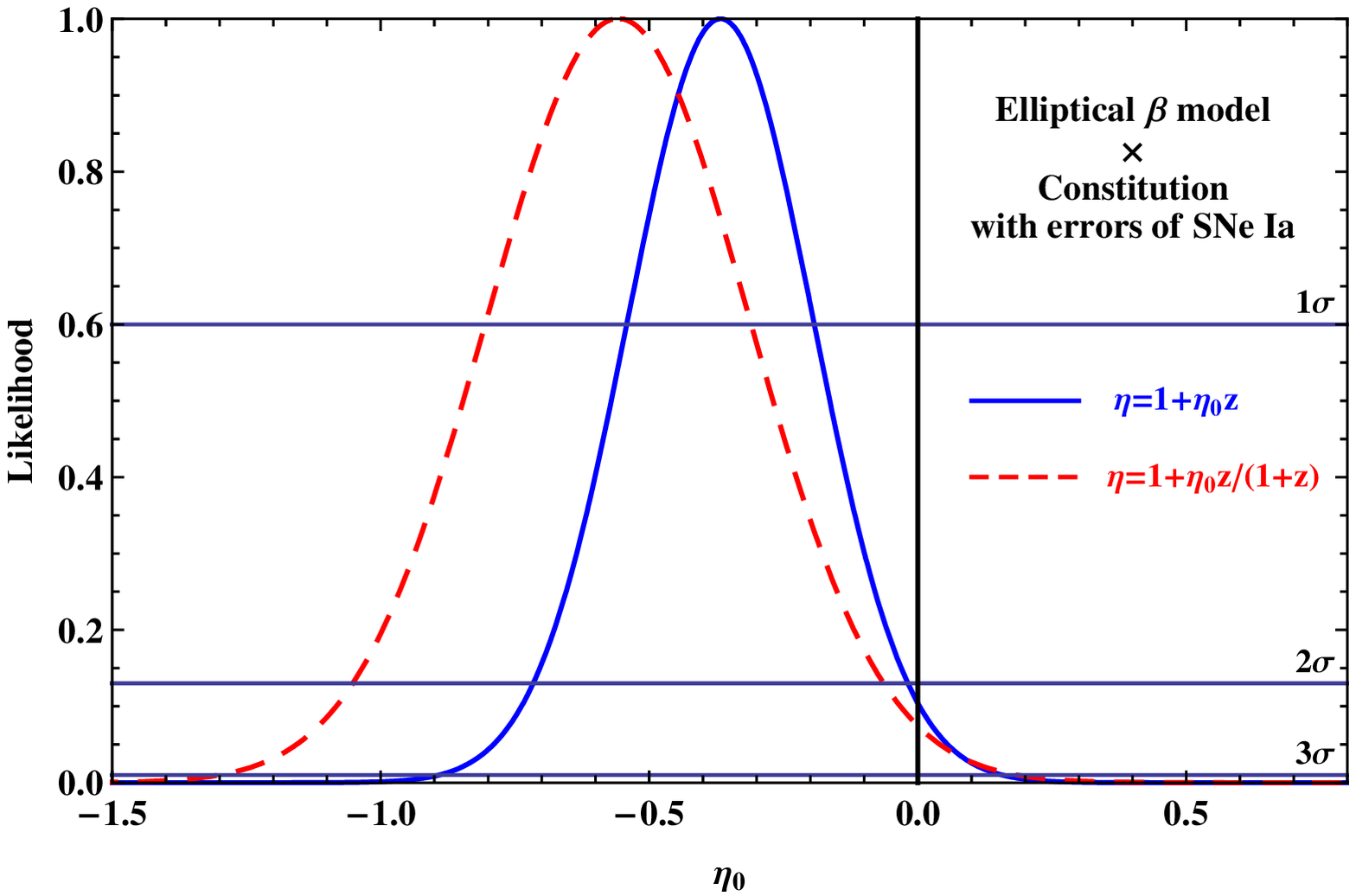}
       \includegraphics[width=0.45\linewidth]{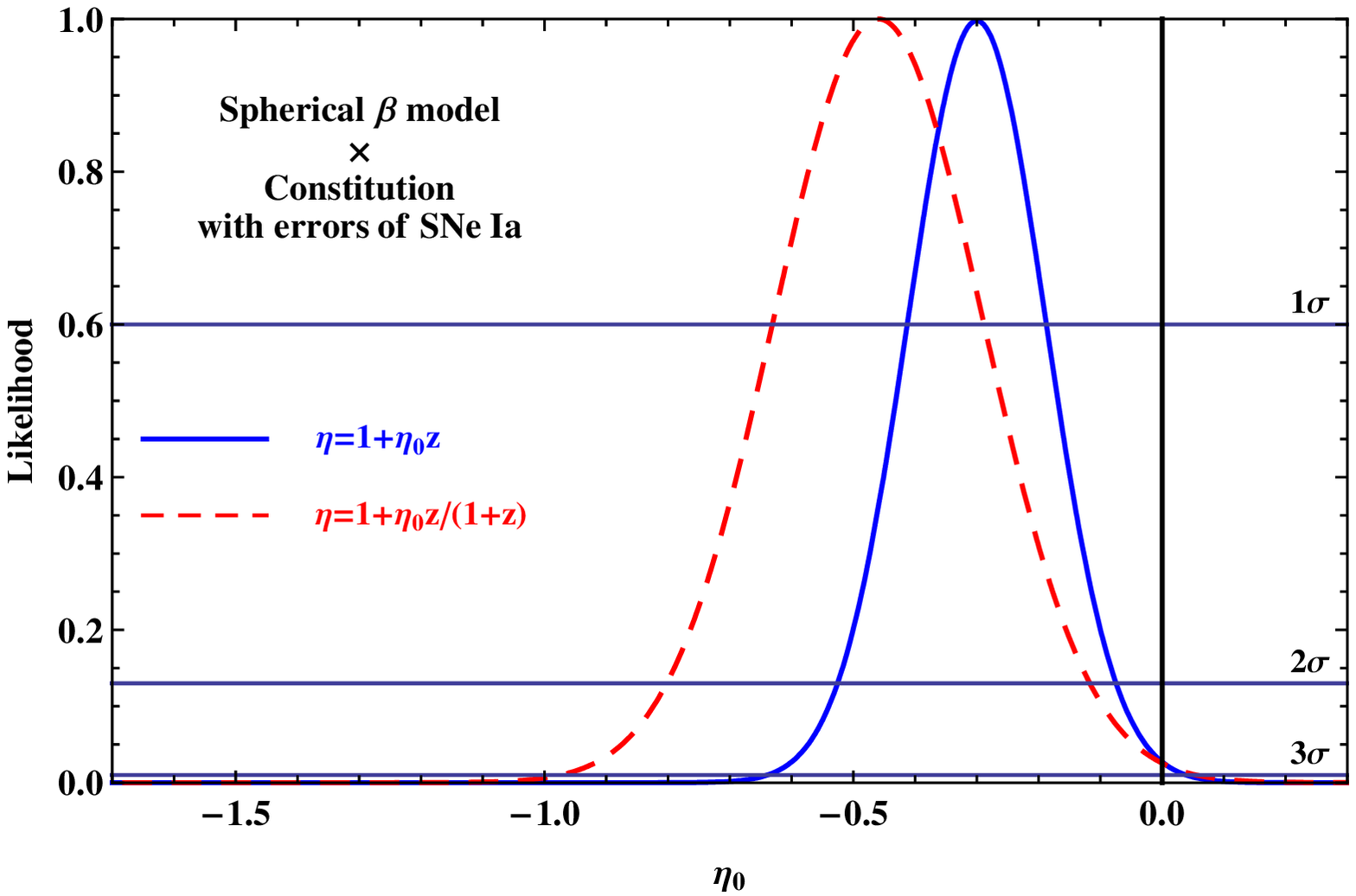}
   \caption{\label{Fig3} $\mathbf{Left}$: likelihood distribution functions from the De Filippis et al.
   (2005) and re-selected Constitution SN Ia pairs for two parameterizations: $\eta(z)=1+\eta_0z$ and
$\eta(z)=1+\eta_0z/(1+z)$. $\mathbf{Right}$: likelihood distribution
functions from the Bonamente et al. (2006)  and Constitution SN Ia
pairs for the same parameterizations. The top and bottom panels
correspond to the cases without and with the errors of SNe Ia,
respectively.}
\end{figure}

We first examine the DD relation by considering two sub-samples
re-selected from Constitution data with two parameterization forms:
$\eta(z)=1+\eta_0z$  and $\eta(z)=1+\eta_0z/(1+z)$. The results are
shown in Fig.~(\ref{Fig3}) and Tab.~(\ref{Tab1}). In this figure,
the top and bottom panels are the results without and with the
errors of SNe Ia, respectively. When the errors of SNe Ia are not
considered, we obtain that, for the De Filippis et al. (2005) sample
(elliptical $\beta$ model), the reciprocity relation is marginally
consistent with observations at $3\sigma$ CL. While the results from
the Bonamente et al. (2006) sample, where a spherical $\beta$ model
was assumed to describe the clusters, show a violation of the DD
relation clearly at $3\sigma$ CL.  Compared with the results by
\citet{Holandab}, where they found that the result from the
elliptical model is compatible with the DD relation at $2\sigma$ CL.
and that from the spherical model is incompatible with it, our
results suggest a stronger violation. For the case with the errors
of SNe Ia taken into account, we find that the results from both the
elliptical and the spherical $\beta$ models  are compatible with the
DD relation at $3\sigma$ CL, which means that the inclusion of the
errors of SNe Ia improves the consistency between the DD relation
and observations.

\begin{figure}[h!]
   \centering
       \includegraphics[width=0.45\linewidth]{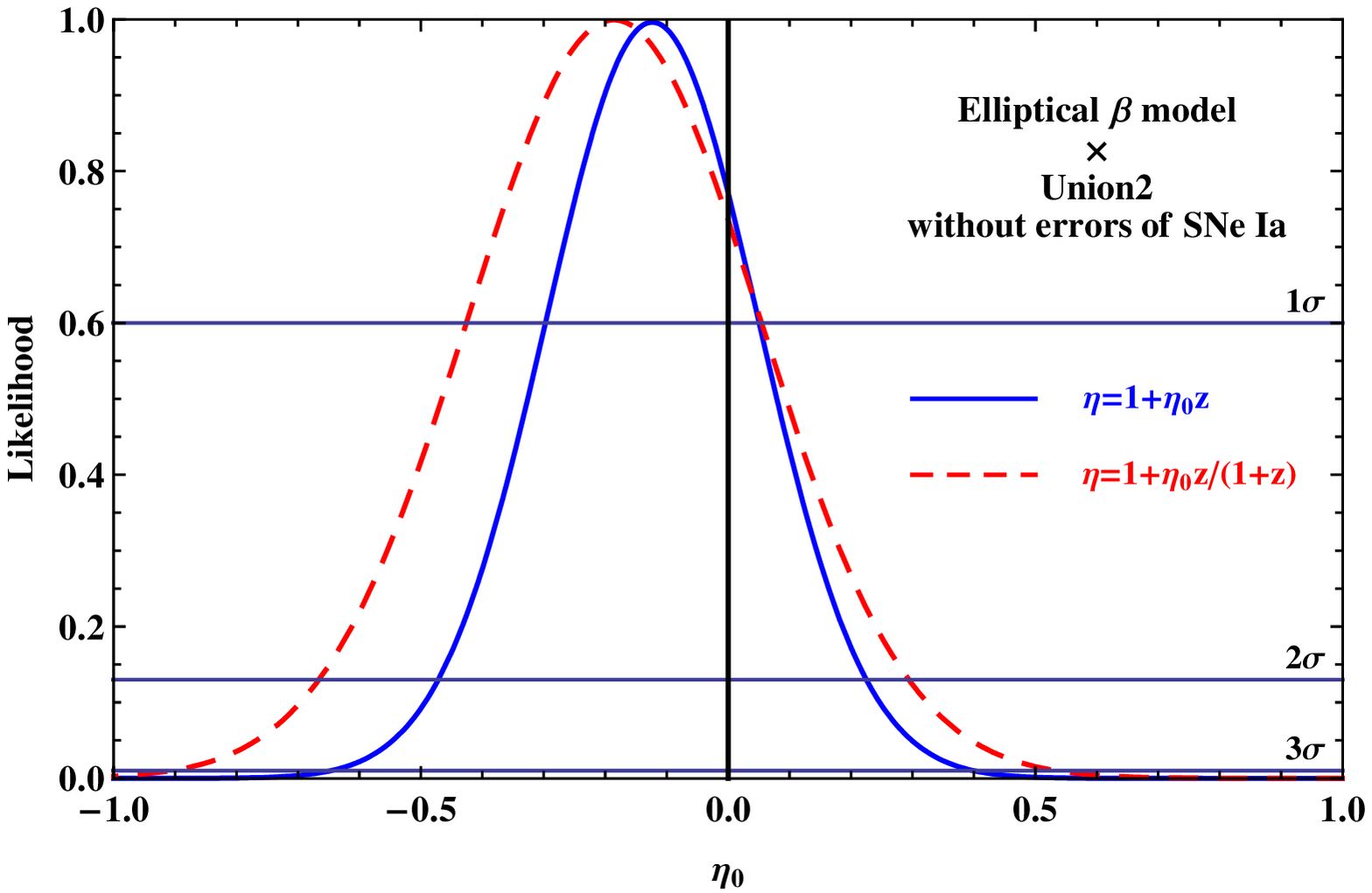}
       \includegraphics[width=0.45\linewidth]{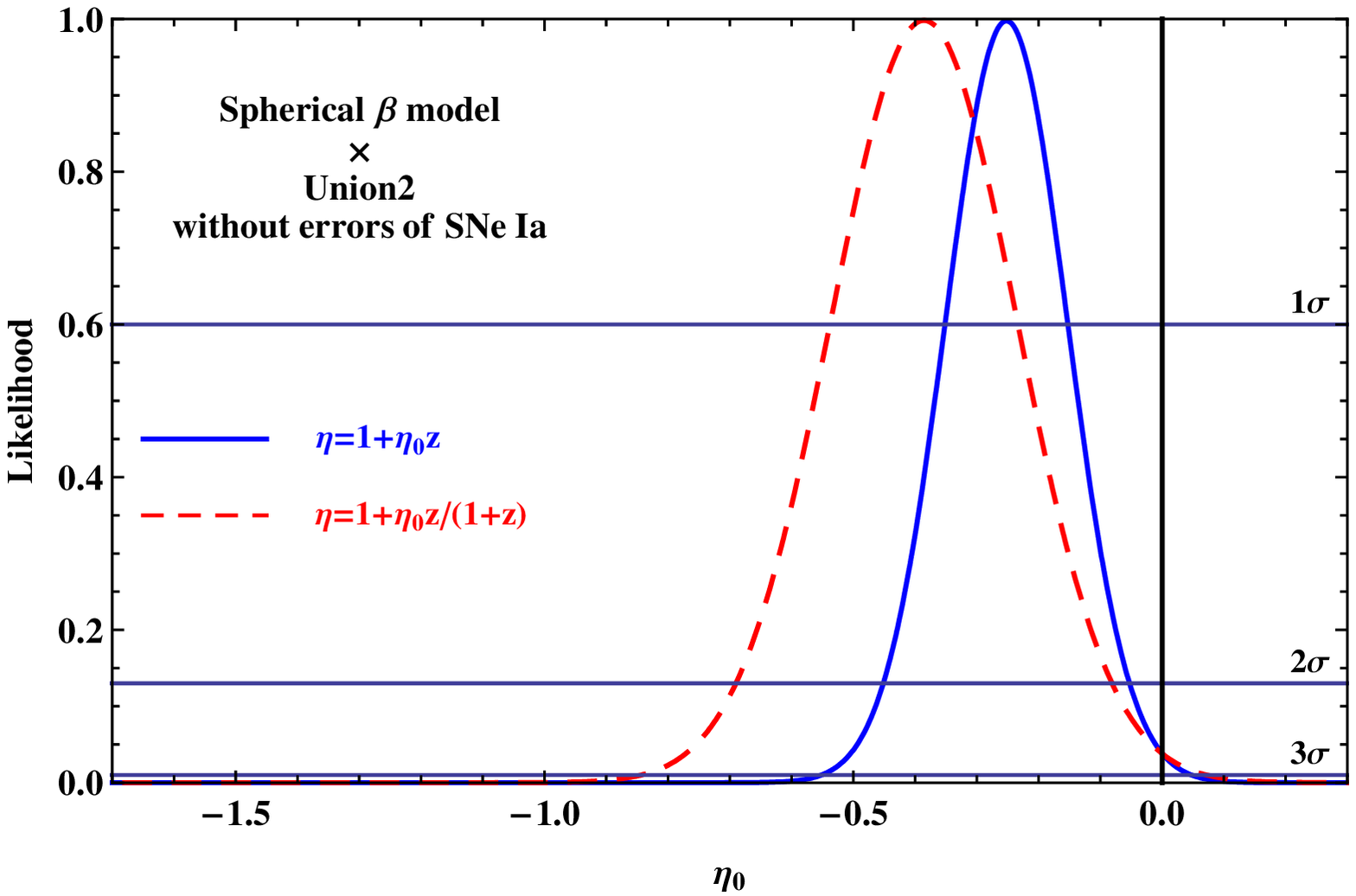}
       \includegraphics[width=0.45\linewidth]{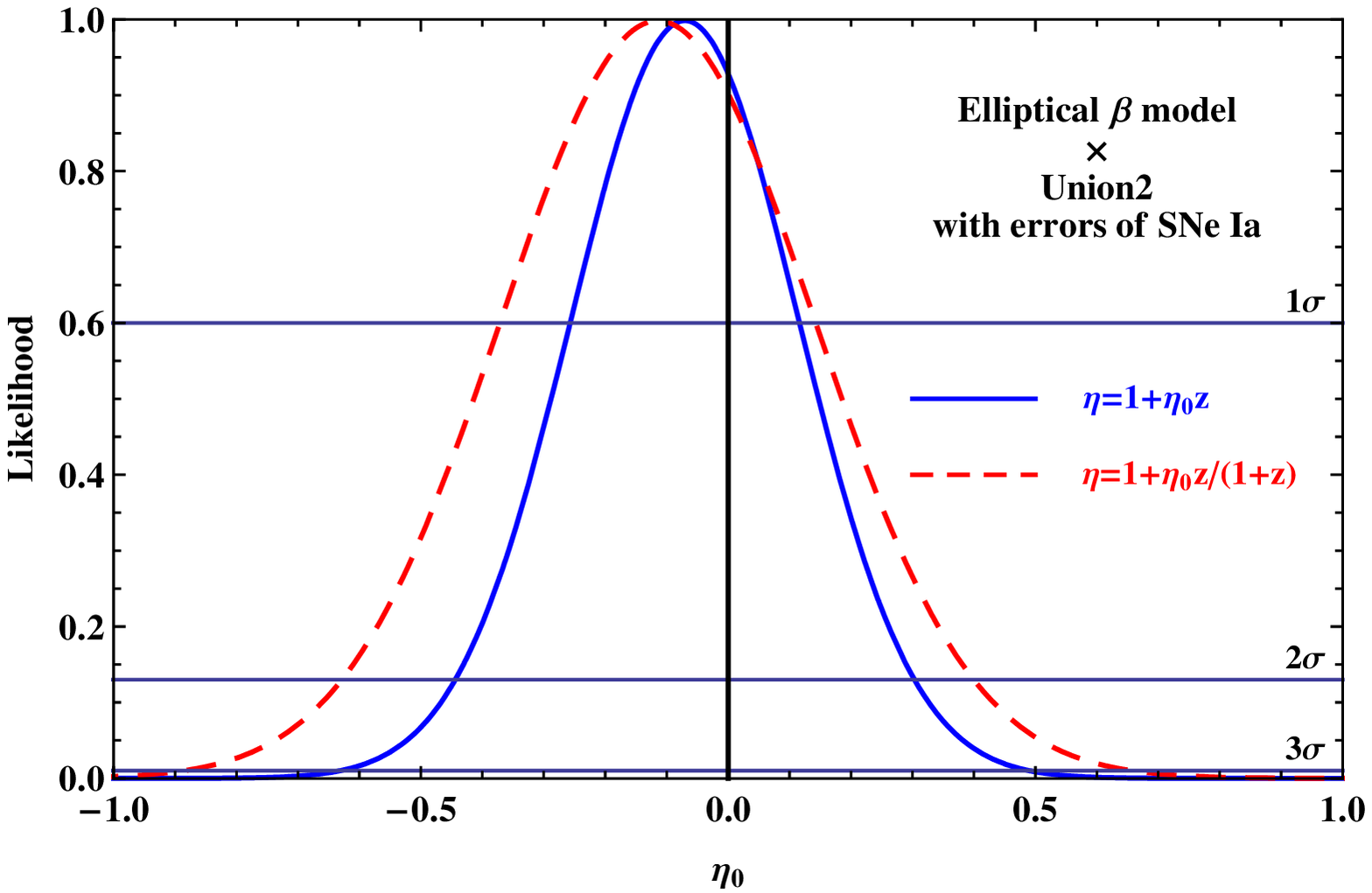}
       \includegraphics[width=0.45\linewidth]{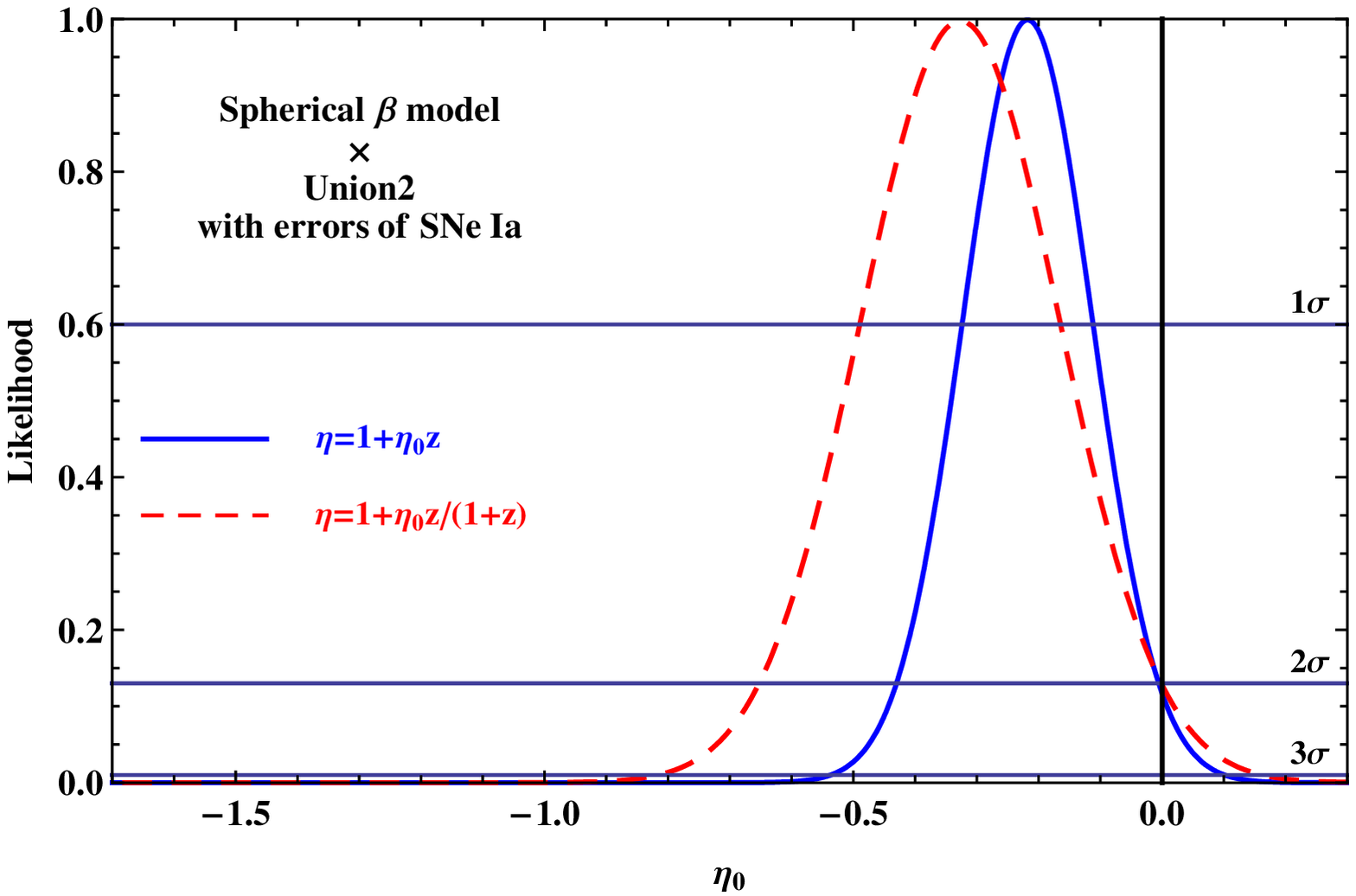}
   \caption{\label{Fig4} $\mathbf{Left}$: likelihood distribution functions from the De Filippis et al.
   (2005)  and Union2 SN Ia pairs for two parameterizations: $\eta(z)=1+\eta_0z$ and
$\eta(z)=1+\eta_0z/(1+z)$. $\mathbf{Right}$: likelihood distribution
functions from the Bonamente et al. (2006)  and Union2 SN Ia pairs
for  the same parameterizations. The top and bottom panels
correspond to the cases  without and with the errors of SNe Ia,
respectively.}
\end{figure}

For  the latest Union2 SNe Ia data,  from the results shown in
Fig.~(\ref{Fig4}) and Tab.~(\ref{Tab1}), we find that,
for both the cases without and with the errors of
SNe Ia, the DD relation can be accommodated at $1\sigma$ C. L. for
the \citet{Filippis} sample and at $3\sigma$ C. L. for the
\citet{Bonamente} sample. The errors of SNe Ia do not affect the
results  significantly, although they tend to make the results  be
more compatible with the DD relation.

\begin{figure}[h!]
   \centering
       \includegraphics[width=0.45\linewidth]{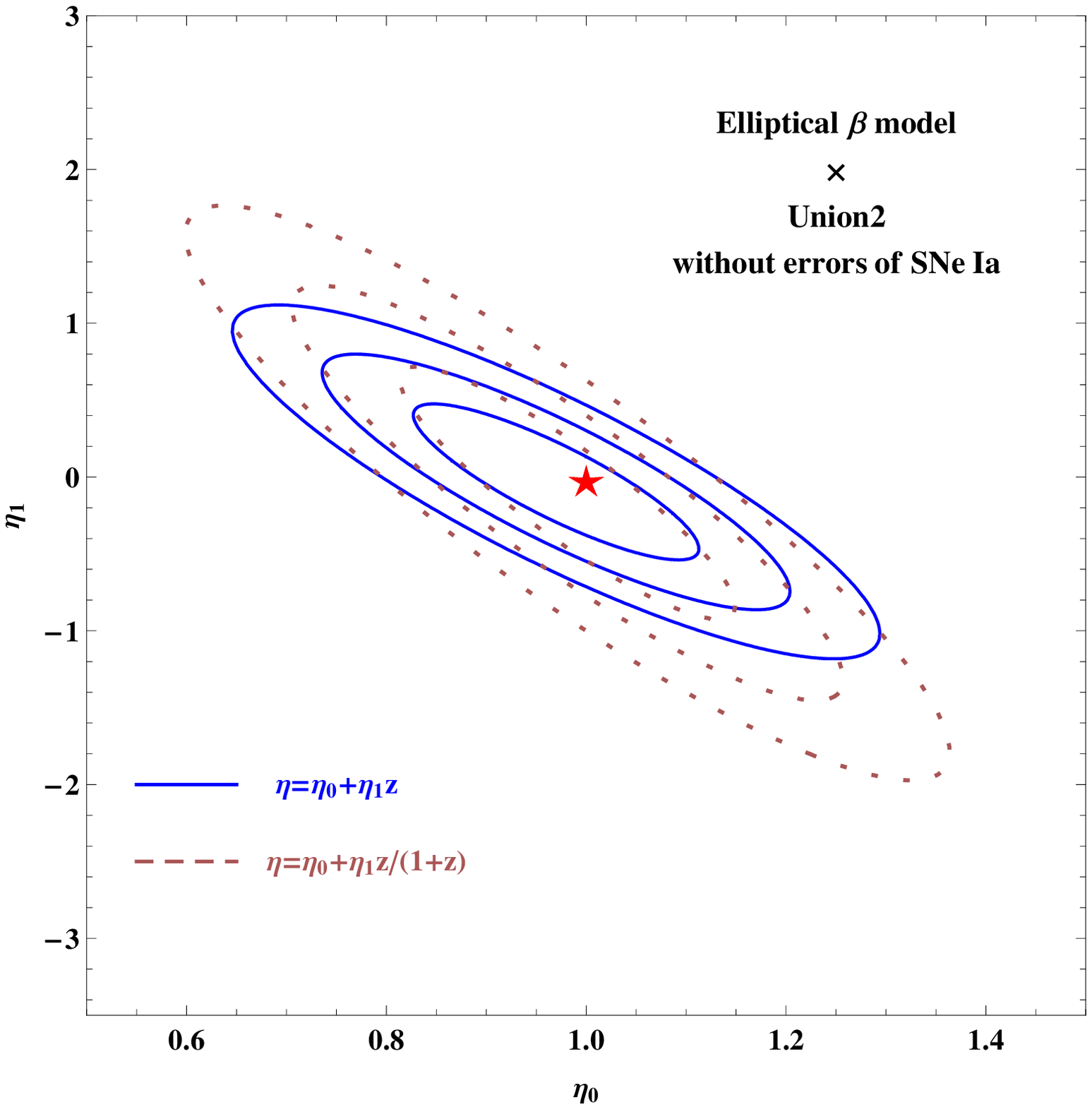}
       \includegraphics[width=0.45\linewidth]{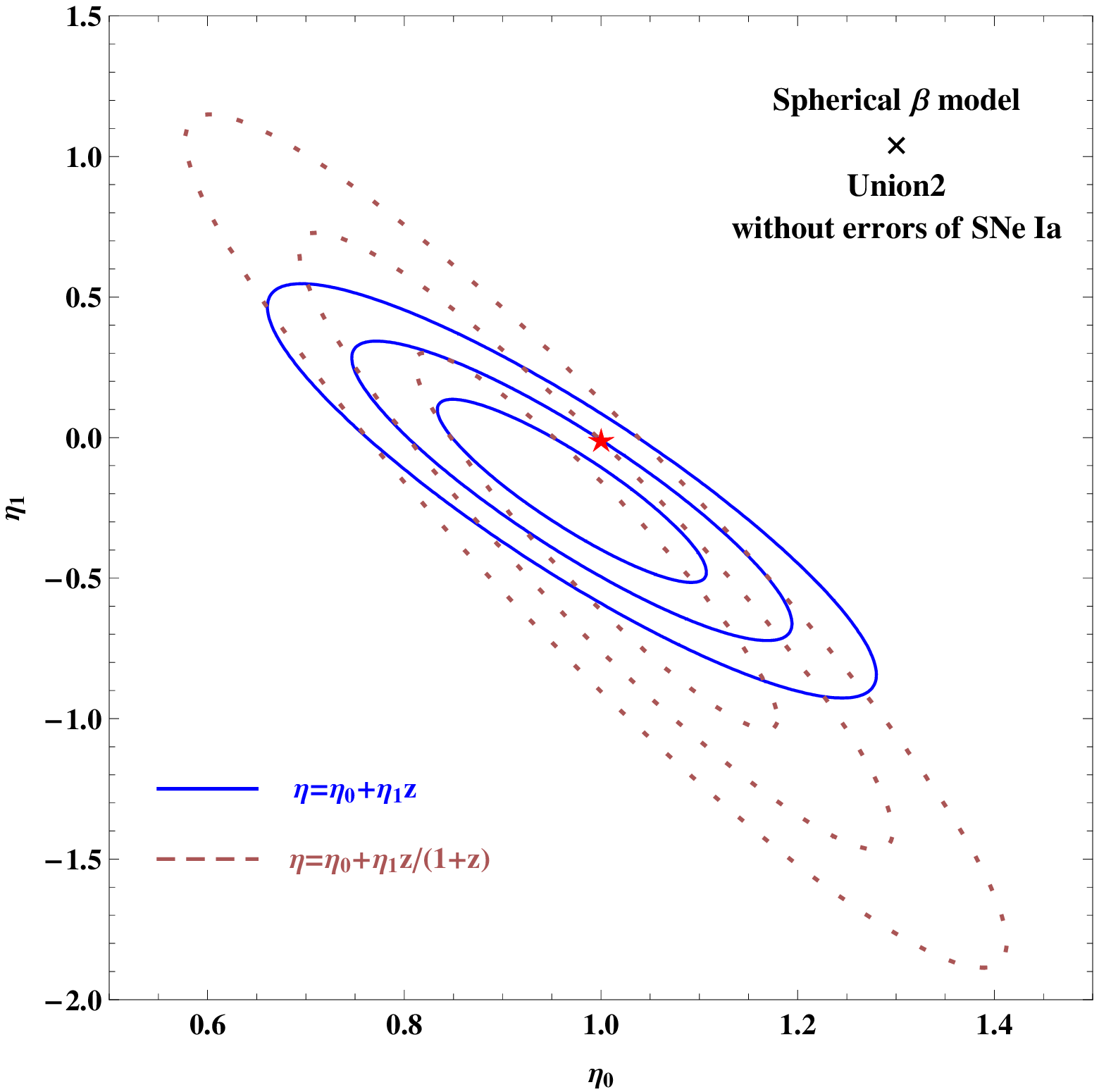}
       \includegraphics[width=0.45\linewidth]{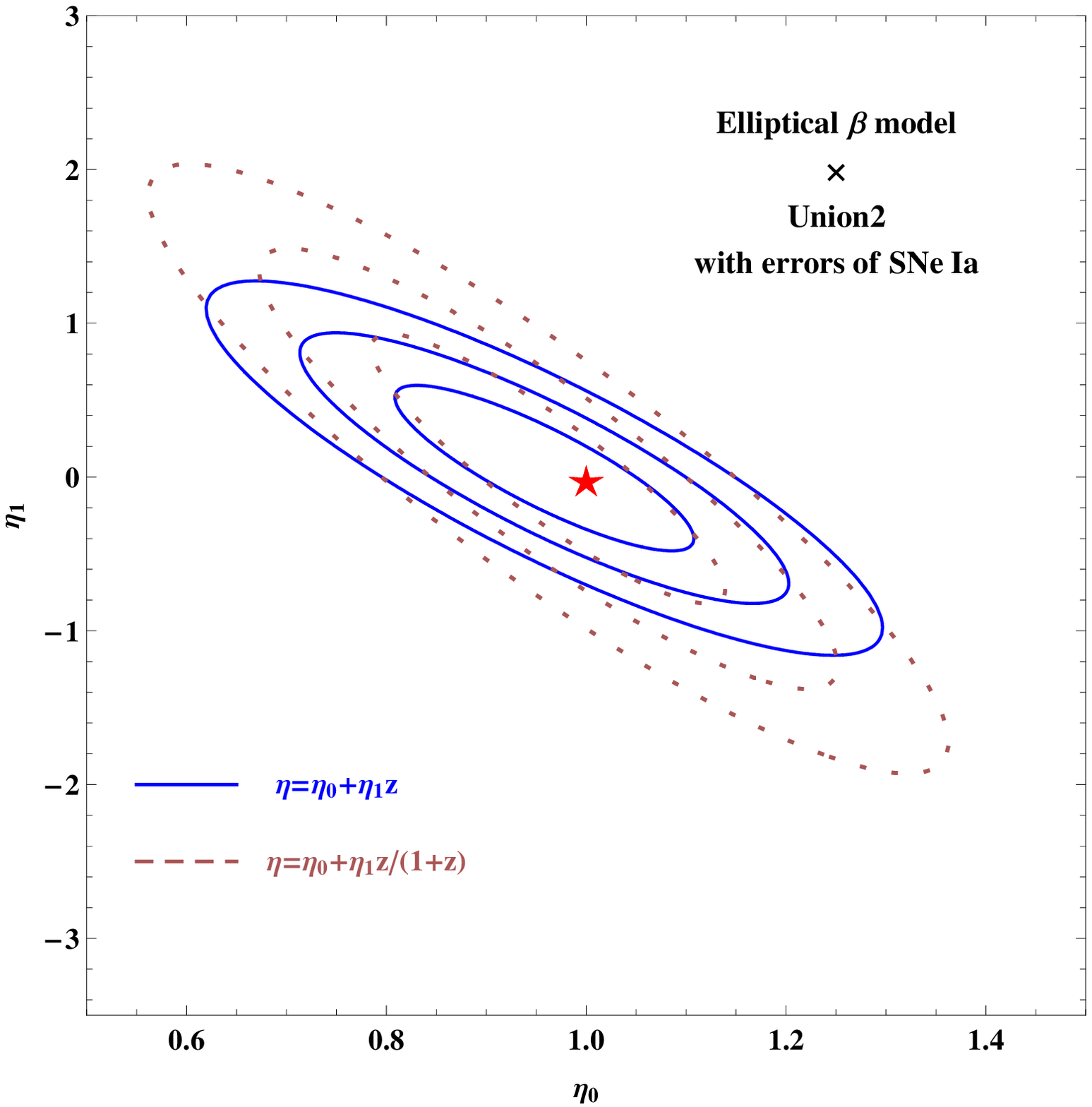}
       \includegraphics[width=0.45\linewidth]{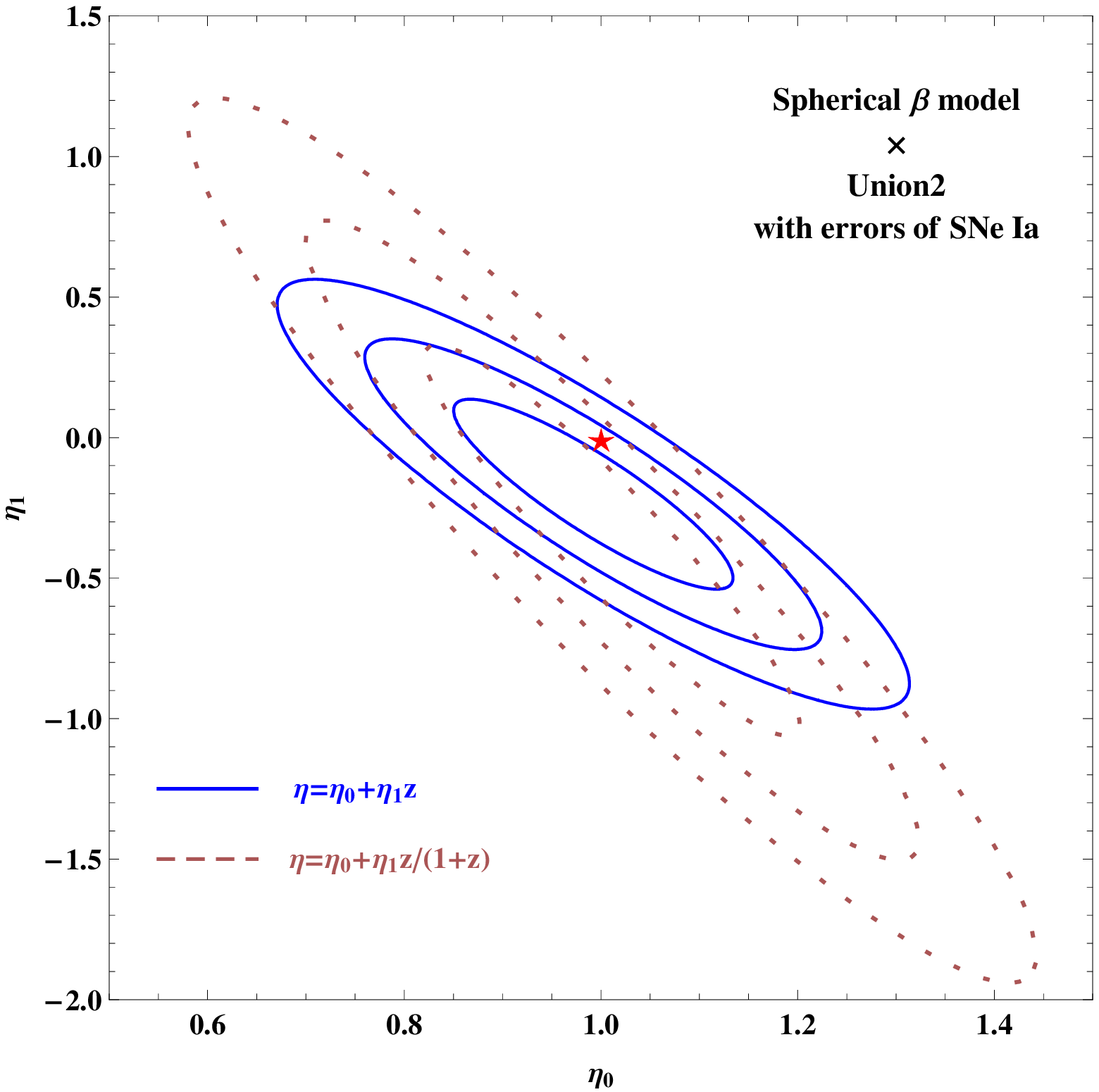}
   \caption{\label{Fig6} $\mathbf{Left}$: likelihood distribution functions from the De Filippis et al.
   (2005) and Union2 SN Ia pairs for two more general parameterizations: $\eta(z)=\eta_0+\eta_1z$ and $\eta(z)=\eta_0+\eta_1z/(1+z)$.
    $\mathbf{Right}$: likelihood distribution functions
    from the Bonamente et al. (2006) and Union2 SN Ia pairs for the same parameterizations. The top and bottom panels correspond to
the cases  without and with the errors of SNe Ia, respectively.}
\end{figure}

Finally,  we also examine the DD relation by assuming two more
general expressions: $\eta(z)=\eta_0+\eta_1z$ and
$\eta(z)=\eta_0+\eta_1z/(1+z)$. The results are shown in
Fig.~(\ref{Fig6}), which  suggest that there is no violation of the
DD relation for the elliptical geometry at $1\sigma$ CL. and for the
spherical $\beta$ model at $2\sigma$ CL.

\begin{table}[!h]\centering
\begin{tabular}{|c|c|c|}
\hline
~ Parameterization(SNe Ia) ~&~~$\eta_0$ (De Filippis et al.) ~~&~~$\eta_0$ (Bonamente et al.) ~~\\
\hline
$1+\eta_0z$ (Constitution) &~~ $-0.40^{+0.17+0.33+0.50}_{-0.17-0.33-0.50}$~~&~~$-0.35^{+0.10+0.20+0.30}_{-0.10-0.20-0.30}$~~\\
\hline
$1+\eta_0z$ (Constitution$^*$) &~~ $-0.37^{+0.18+0.35+0.52}_{-0.18-0.35-0.52}$~~&~~$-0.30^{+0.11+0.23+0.34}_{-0.11-0.23-0.34}$~~\\
\hline
$1+\eta_0\frac{z}{1+z}$ (Constitution) &~~ $-0.60^{+0.24+0.47+0.70}_{-0.24-0.47-0.70}$~~&~~$-0.54^{+0.15+0.31+0.46}_{-0.15-0.31-0.46}$~~\\
\hline
$1+\eta_0\frac{z}{1+z}$ (Constitution$^*$) &~~ $-0.56^{+0.25+0.49+0.74}_{-0.25-0.49-0.74}$~~&~~$-0.46^{+0.17+0.34+0.51}_{-0.17-0.34-0.51}$~~\\
\hline
$1+\eta_0z$ (Union2) &~~ $-0.12^{+0.17+0.35+0.52}_{-0.17-0.35-0.52}$~~&~~$-0.25^{+0.10+0.20+0.30}_{-0.10-0.20-0.30}$~~\\
\hline
$1+\eta_0z$ (Union2$^*$) &~~ $-0.07^{+0.19+0.37+0.56}_{-0.19-0.37-0.56}$~~&~~$-0.22^{+0.11+0.21+0.32}_{-0.11-0.21-0.32}$~~\\
\hline
$1+\eta_0\frac{z}{1+z}$ (Union2) &~~ $-0.19^{+0.24+0.50+0.74}_{-0.24-0.50-0.74}$~~&~~$-0.39^{+0.15+0.31+0.46}_{-0.15-0.31-0.46}$~~\\
\hline
$1+\eta_0\frac{z}{1+z}$ (Union2$^*$) &~~ $-0.11^{+0.26+0.51+0.77}_{-0.26-0.51-0.77}$~~&~~$-0.33^{+0.16+0.33+0.49}_{-0.16-0.33-0.49}$~~\\
\hline
\end{tabular}
\tabcolsep 0pt \caption{\label{Tab1} Summary of the Results for
$\eta(z)=1+\eta_0 z$ and $\eta(z)=1+\eta_0\frac{z}{1+z}$,
respectively, at $1,2$, and $3\sigma$ Confidence Levels for the
Cases without and with the Errors of SNe Ia. In the table, an
asterisk represents the case with the errors of SNe Ia considered.}
\vspace*{5pt}
\end{table}

\section{CONCLUSION}
In this Letter,  we test the DD relation by considering the ADDs
given by two samples of galaxy clusters together with the luminosity
distances provided by sub-samples of SNe Ia picked from the
Constitution and the latest Union2 data sets. The Constitution
sample has already been discussed by \citet{Holandab}, and they
found that for both ADD samples three data points should be removed
with the selection criteria ($\Delta z=
\left|z_{Cluster}-z_{SNe~Ia}\right|<0.005$). However, we find that,
with the same selection criteria, the data points that have to be
removed are actually  six and twelve respectively for  the
\citet{Filippis} sample and the \citet{Bonamente} sample. A
re-analysis with more data points discarded suggests a violation of
the DD relation stronger than that given in \citet{Holandab}. In
order to obtain a more reliable result, we investigate the DD
relation by considering the latest Union2 SNe Ia. It is worthy to
note that with the Union2 SNe Ia all ADD data can be retained  and
the differences of the redshifts between ADD from the galaxy cluster
and the associated luminosity distance from SNe Ia are much smaller.
Thus the accuracy of our test should be improved. Our results then
show that the DD relation can be accommodated at $1\sigma$ CL. for
the elliptical $\beta$ model \citep{Filippis} and  at $3\sigma$ C.
L. for the spherical $\beta$ model \citep{Bonamente}. Finally, we
examine the DD relation by postulating two more general
parameterization forms: $\eta(z)=\eta_0+\eta_1z$ and
$\eta(z)=\eta_0+\eta_1z/(1+z)$, and we find  that the consistencies
between the observations and the DD relation are improved markedly
for both samples of galaxy clusters. The DD relation is compatible
with \citet{Filippis} sample and \citet{Bonamente} sample at
$1\sigma$ and $2\sigma$ CL., respectively.  Furthermore, with the
inclusion of the errors of SNe Ia, the results become more
consistent with the DD relation. Therefore, our results suggest that
the DD relation is compatible with the observations. This differs
from what is obtained by \citet{Holandab}, where the results from
the \citet{Bonamente} sample give a clear violation of the DD
relation.

\acknowledgments This work was supported in part by the National
Natural Science Foundation of China under Grants Nos.  10935013 and
11075083, Zhejiang Provincial Natural Science Foundation of China
under Grant No. Z6100077, the FANEDD under Grant No. 200922, the
National Basic Research Program of China under Grant No.
2010CB832803, the NCET under Grant No. 09-0144, and the PCSIRT under
Grant No. IRT0964.

\end{document}